\documentclass[conference]{IEEEtran}
\IEEEoverridecommandlockouts

\usepackage{cite}
\usepackage{amsmath,amssymb,amsfonts}
\usepackage{algorithmic}
\usepackage{graphicx}
\usepackage{textcomp}
\usepackage{xcolor}
\usepackage{xspace}
\usepackage{todonotes}
\usepackage{hyperref}

\graphicspath{{figures/}{pictures/}{images/}{./}} 

\def\BibTeX{{\rm B\kern-.05em{\sc i\kern-.025em b}\kern-.08em
    T\kern-.1667em\lower.7ex\hbox{E}\kern-.125emX}}
\makeatletter
\newcommand*\titleheader[1]{\gdef\@titleheader{#1}}
\AtBeginDocument{%
  \let\st@red@title\@title
  \def\@title{%
    \bgroup\normalfont\large\centering\@titleheader\par\egroup
    \vskip1.5em\st@red@title}
}
\makeatother

\title{
Immersive Analytics of Large Dynamic Networks via Overview and Detail Navigation
}
\titleheader{\copyright 2019 IEEE. DOI: 10.1109/AIVR46125.2019.00030}

\begin{document}



\newcommand{\sys}{\textit{System Name}\xspace}


\author{\IEEEauthorblockN{Johannes Sorger}
\IEEEauthorblockA{\textit{Complexity Science Hub} \\
Vienna, Austria \\
sorger@csh.ac.at}
\and
\IEEEauthorblockN{Manuela Waldner}
\IEEEauthorblockA{\textit{TU Wien} \\
Vienna, Austria \\
waldner@cg.tuwien.ac.at}
\and
\IEEEauthorblockN{Wolfgang Knecht}
\IEEEauthorblockA{\textit{Complexity Science Hub} \\
Vienna, Austria \\
knecht@csh.ac.at}
\and
\IEEEauthorblockN{Alessio Arleo}
\IEEEauthorblockA{\textit{TU Wien} \\
Vienna, Austria \\
alessio.arleo@tuwien.ac.at}
}

\maketitle

\begin{abstract}

Analysis of large dynamic networks is a thriving research field, typically relying on 2D graph representations. The advent of affordable head mounted displays however, sparked new interest in the potential of 3D visualization for immersive network analytics.
Nevertheless, most solutions do not scale well with the number of nodes and edges and rely on conventional fly- or walk-through navigation.
In this paper, we present a novel approach for the exploration of large dynamic graphs in virtual reality that interweaves two navigation metaphors: overview exploration and immersive detail analysis.
We thereby use the potential of state-of-the-art VR headsets, coupled with a web-based 3D rendering engine that supports heterogeneous input modalities to enable ad-hoc immersive network analytics.
We validate our approach through a performance evaluation and a case study with experts analyzing a co-morbidity network.
\end{abstract}


\section{Introduction}\label{se:intro}

Virtual and augmented reality (VR and AR) are by no means new concepts: already in 1968, Sutherland presented a first, three-dimensional head mounted display~\cite{sutherland1968head}. In 1993, thanks to the technical advancements of 3D hardware, the IEEE Conference on Virtual Reality (IEEE VR) was held for the first time. As such, research on graph visualization in virtual reality has already been conducted in the 90ies \cite{crutcher1995managing}.
Since then, several studies have shown that stereoscopy can improve the users' understanding of graphs as compared to monoscopic 3D and 2D graph drawing \cite{ware1996evaluating, ware2008visualizing, belcher2003using, greffard2011visual, halpin2008exploring, kwon2016study, marriott2018immersive_chapter}. Raja et al.~\cite{raja2004exploring} showed that increasing immersion in a CAVE, by supplying more display space and supporting head tracking, had a positive effect on the users' analysis of 3D scatterplots.
In comparison, VR experienced through a head-mounted display (HMD) is even more effective than the immersive environment of CAVE-like solutions~\cite{demiralp2006cave}. For instance, users could collaboratively analyze graph connectivity significantly faster using HMDs than in a CAVE-like environment \cite{cordeil2016immersive}. Similar benefits have been shown for 3D scatterplots: an immersive VR setup could increase classification accuracy and led to higher engagement by the users compared to 2D and 3D desktop rendering \cite{wagner2017immersive, wagner2018immersive}. Bowman and McMahan \cite{bowman2007virtual} argue that increased immersion can reduce visual clutter and strengthen the comprehension of the displayed scene.
Despite these encouraging findings and the increasing availability of commercial HMDs, many visualization researchers kept a conservative attitude towards immersive environments and their potential data-mining applications~\cite{marriott2018immersive}. In recent network visualizations, the use of 3D techniques is still rare, and little research has investigated novel graph drawing or interaction techniques for immersive network analytics.

Considering these promising previous results and the availability of affordable state-of-the-art VR headset solutions, in this paper we investigate the challenge of enabling immersive analytics for large dynamic networks using modern VR headsets.
We explore the possibilities of ``immersion'' into the network for detailed analysis while being provided with the context of a well-known overview perspective on the network to stay oriented.
Our proposed exploration technique takes advantage of modern immersive technologies using a flexible, expandable and browser-based software infrastructure natively compatible with VR headsets but also supporting traditional input methods to support ad-hoc immersive analytics also for users without programming experience. Our contributions are listed in the following:

\begin{itemize}
    \item A novel approach for the immersive exploration of networks in VR via overview\&detail navigation;
    \item A fast web-based render engine for large dynamic graphs offering VR controller support for ad-hoc immersive network analytics;
    \item First encouraging results of a case study with domain experts analyzing medical data (a co-morbidity network), showing the potential of VR for network analytics.
\end{itemize}



\begin{figure*}[t!]
    \centering
    \includegraphics[width=\linewidth]{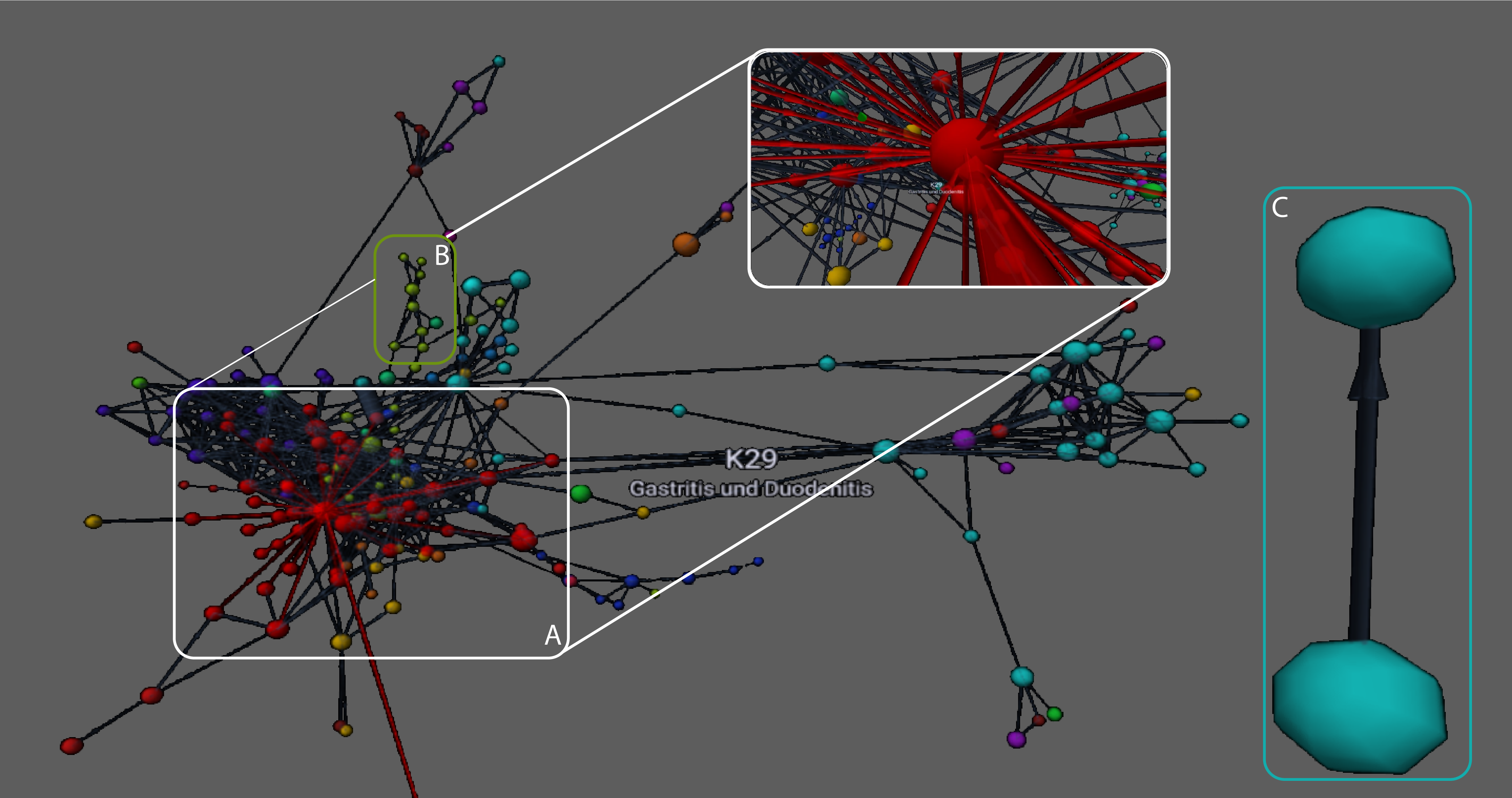}
    \caption{The overview perspective of a co-morbidity network. A) K29, a ``hub'' node with a high degree; the inset shows the same node from the detail perspective. The node is selected, therefore its neighbors and out-edges are highlighted in red and the node label is displayed.  B) A cluster. C) An isolated component of the graph. (For better readability on printed paper the background color and contrast of the image have been adapted.)
    }
    \label{fig:overview}
\end{figure*}

\section{Related Work}\label{se:rel}

Adapting visualizations to new modalities, such as VR, is not a trivial task, as it may or may not require new ways to represent and interact with the data \cite{sadana2016redefining}. Researchers therefore have explored ways of how to interactively \emph{manipulate}, \emph{navigate}, and \emph{represent} networks in VR. Already in an early CAVE-like environment, Osawa et al.~\cite{osawa2000immersive} introduced hand gestures to manipulate distant nodes of a graph. They also introduced a variation of a ``spotlight'' \cite{forsberg1996aperture} that distorts the graph around the focus region for excocentric navigation around the graph. More recently, Huang et al.~\cite{huang2017gesture} as well as Erra et al.~\cite{erra2019virtual} presented a natural graph exploration and manipulation interface for VR using a Leap Motion sensor. Drogemuller et al.~\cite{drogemuller2017vrige} introduced a virtual ``filter cube'' to filter and highlight nodes in the graph.

Besides manipulation, navigation is considered to be one of the key aspects to be solved for information visualization in 3D \cite{brath20143d}. Modern graph visualization in HMD-based VR usually provides a fly-through interface using a 6 DOF camera control \cite{wagner2018immersive,erra2019virtual} or the ability to walk through the graph \cite{drogemuller2017vrige}.
Drogemuller et al.~\cite{drogemuller2018evaluating} compared four VR navigation techniques for graph exploration using room-sized VR systems. They found that two-handed flying \cite{mine1997moving} using two controllers was most efficient for finding single nodes or paths between two nodes in the graph.
While this work compared the appropriateness of established VR navigation techniques, our goal was to explore a new navigation method to facilitate immersive network analytics in particular.

Most prior immersive network analytics systems show networks as conventional node-link diagrams based on a force-directed 3D graph layout. Some also experimented with other ways of how to represent networks in VR or AR.
B\"uschel et al.~\cite{buschel2019augmented} evaluated different edge styles for AR graphs and found that curved edges are harder to interpret. Halpin et al.~\cite{halpin2008exploring} initialized their VR scene with a 2D graph layout. Users could interactively ``extrude'' nodes to the third dimension to explore the social network.
Kwon et al.~\cite{kwon2015spherical} suggested to render graphs using a spherical layout in VR, optimized towards usage while seated. Interaction was carried out via mouse and keyboard.
Compared to a 2D representation of the graph in VR, immersive visualization allowed users to successfully complete more complex tasks on larger graphs than before \cite{kwon2016study}.
Marriot et al.~\cite{marriott2018immersive_chapter} argue that egocentric data views, such as proposed by Kwon et al.~\cite{kwon2015spherical, kwon2016study}, may increase user engagement. However, a study by Yang et al.~\cite{yang2018maps} showed that maps explored from an egocentric perspective led to lower performance than any other representation in VR (an exocentric view, a flat map view, and a curved map view). We share the belief that egocentric data views are a unique strength of virtual environments to provide a new perspective on the data that cannot be easily obtained on a desktop screen. In contrast to Kwon et al.~\cite{kwon2015spherical, kwon2016study}, however, we also believe that a conventional exocentric view on a network provides a visual anchor for orientation and overview. In contrast to these prior works, we therefore present and explore a new immersive network analytics approach, where users can flexibly switch between an exocentric overview and an egocentric detail view.


\section{Design and Requirements}\label{se:design}

Our goal is to provide new insights in the course of network analytics, such as social network analytics \cite{aggarwal2011introduction} or network medicine \cite{barabasi2011network}, by providing a new visual perspective of the data, enabled through VR.
To design our system, we make use of the ``Data-Users-Tasks'' design triangle by Miksch and Aigner~\cite{miksch2014matter}. This technique requires the definition of three crucial elements that will steer the design process: the \textbf{Data} the system will process, the \textbf{Users} that will get in contact with the system, and the \textbf{Tasks} the system is expected to perform. In this section, we will discuss each of these aspects. 

\subsection{Data}\label{se:design_data}

To face the challenges of modern network analytics tasks \cite{aggarwal2011introduction, barabasi2011network}, our system is expected to handle large, dense, dynamic network data.
Nodes and edges can be associated with additional attributes, such as a cluster membership for nodes or edge weights that rank a connection among the others. In the scope of this paper, we assume temporal changes of the graph to be known beforehand, and to be discrete, i.e., they are grouped into distinct time ``frames''.

\subsection{Users}\label{se:design_users}

Large network exploration is a difficult task to perform, in 2D and 3D alike. Our target users are researchers who wish to explore large networked domain specific data using an intuitive and life-like approach, in which they can manipulate, move around, and interact with the data as if it would be physically present with them. 
We aim to support users who are not necessarily experts or knowledgeable in network visualization, graph drawing, or even data science and visualization, i.e., from disciplines such as healthcare, finance, or social science.

\subsection{Tasks}\label{se:design_tasks}

The purpose of our system is to provide an immersive exploration and interaction technique to obtain (expected and unexpected) insights from dynamic networked data.
Typical network analytics tasks across domains have been identified and summarized by Lee et al.~\cite{lee2006task}. Driven by the challenges of modern network analytics approaches in different domains \cite{aggarwal2011introduction, barabasi2011network}, we focus here mainly on \emph{topology-based tasks} \cite{lee2006task}, including:
\begin{itemize}
    \item finding \emph{hubs}, such as highly connected nodes in a co-morbidity network or the most influential members of a social network,
    \item detection of \emph{communities}, such as tight interaction between proteins or highly connected social communities,
    \item locating the shortest molecular \emph{paths}, or
    \item characterizing the \emph{dynamics} of social interaction.
\end{itemize}
In addition, we speculate that it is important to first obtain an \emph{overview} of the network, as stipulated in the widely accepted information seeking mantra \cite{shneiderman1996eyes}. Gaining an overview is essential to infer global insights out of the network's topology, allowing users to identify interesting spots and areas suitable for more in-depth evaluation. Finally, the system should provide \emph{details-on-demand} to inspect attributes of single nodes and edges to derive detailed information from local structures.

We believe that immersive network analytics is an appropriate method for supporting topology-based exploration and browsing through large, dense graphs as it can provide a new, complementary perspective of the data. Using VR, the user can inspect parts of the network in detail, i.e., from a first person perspective, e.g., by following a path of edges through the network, or by inspecting local connectivity from an angle with better visibility; thus enabling browsing and topology-based tasks~\cite{lee2006task}.

\section{System}\label{se:sys}

In this section, we describe the system in detail, both in terms of visual design and interactions.



\subsection{Network Visualization}

Our goal is to provide expert users with a new perspective on their data in VR. It is therefore important to show the data as closely as possible to their conventional representation form to support quick orientation. Hence, we render the network using the most wide-spread graph rendering method: node-link diagrams. Nodes are rendered as spheres of different diameters. Edges are represented as tubes. In case of a directed graph, the direction of each edge is represented by an arrow, close to and directed towards the target node (see \autoref{fig:overview}C). 
The color and diameter of the nodes as well as the color and girth of the edges can be changed according to user-defined attribute values in the graph input file.
The 3D layout of the graph is delegated to a 3D extension of the D3 force-layout ~\cite{d3force, d3library}.


\subsection{User Perspectives}

A special design consideration of our system is the notion to let users explore the network and its evolution from two different ``perspectives'': \textbf{overview} and \textbf{detail}. In the overview, users regard the graph from ``outside'', meaning the graph is positioned so that it completely fits within the user's view. This enables the user to look at the entire network (from different angles), in order to get an overview of interesting node/edge formations and clusters -- i.e., to spot interesting parts of the graph that they would like to explore in more detail. The overview thereby shares similarities to a conventional 2D graph layout, which expert users analyzing networks are usually very familiar with. 

If the users want to inspect a part of the network in more detail, they can select a specific node to immerse themselves into the detail perspective. From this perspective, users can effectively explore local neighborhoods even in very dense networks.


After teleportation between perspectives, users may have difficulties staying oriented. To improve orientation, the previous point of view is therefore represented within the context of the graph as a camera shaped object, informing the user on the position from where the exploration started.
To further aid orientation and to facilitate switching perspectives, in the detail perspective, the overview camera's position is indicated as a green arrow in the top of the user's view.
The arrow guides the user towards the overview camera's position by pointing to its direction (see Fig.\ref{fig:time}).



\subsection{Navigation and Interaction}\label{se:sys_interaction}

\textbf{Input: }In the current version of our system, we designed interaction techniques specifically for the HTC Vive. However, the controls can be adapted to any input device that features the same set of input modalities. The HTC Vive (akin to other state of the art solutions) features one wireless controller for each hand, to allow users to use their hands freely and independently from each other - as opposed to conventional game-pads that are held with both hands together.
Since we aimed to make interaction with the graph intuitive also for users who are not yet familiar with VR technology, the system is designed to work with only one controller as input device. The user can thus single-handedly interact with and navigate within the graph. In the same vein, the controls do not need to be specifically adjusted for left- or right-hand users.

\begin{figure}[t]
    \centering
    \includegraphics[width=\linewidth]{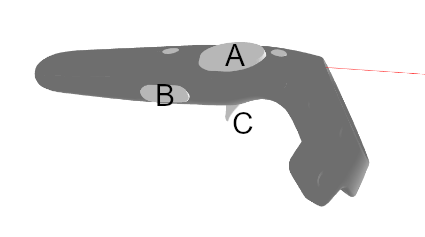}
    \caption{A depiction of the virtual representation of the controller: A)~Directional pad; B)~Side button for D-pad input-modification (temporal navigation); C)~Trigger for node/edge selection. The vertical axis of the controller determines the laser pointer orientation.
    }
    \label{fig:controller_combo}
\end{figure}

Required input modalities (as supported by the HTC Vive controller) are the position and orientation of the controller, a two-dimensional track-pad enabling input on an x- and y-axis (i.e., directional pad, or D-pad), two additional buttons (a trigger button and an input modifier button).
Further, the VR headset position and rotation have to be traceable. 
The button layout on the HTC Vive controller is depicted in Figure \ref{fig:controller_combo}.

In general, input can be differentiated into graph interactions and graph navigation.

\textbf{Graph interactions: }Graph interactions are carried out with the ``laser-pointer'', i.e., a visible ray that is cast from the controller into the scene, according to the controller's position and orientation (see Fig.\ref{fig:controller_combo}). Upon hovering with the laser-pointer on a graph element, the user can inspect details
, such as a node label or an edge weight.
Such details on node or edge attributes are displayed as a text label in the center of the user's view (see Fig.\ref{fig:overview}, center)
, in order achieve the best readability.
Due to the curvature of the magnifying lenses in the headset,
text in the periphery can be distorted and blurred, and would thus be hard to read.
Hovering on a node also reveals additional details of the network, by allowing a user to inspect the direct neighborhood 
of the hovered node (see Fig.\ref{fig:overview}-A). Direct neighbors of the hovered node are thereby highlighted in red, while all other graph elements that are not part of this neighborhood are lowlighted so that they blend into the background . This is achieved by reducing their opacity and giving them a uniform dark hue.
Graph interactions function the same way independent of the chosen perspective.

\textbf{Graph navigation: }Graph navigation on the other hand, offers a different navigation method for each perspective (i.e., overview and detail) that is designed to complement the respective task. 
In both modes, the D-pad can be used for direct navigation.
In the overview, input on the D-pad rotates the graph around its local axes, enabling the user to quickly get an impression of the network from various angles. The entire graph is thereby enclosed within a (non-visible) bounding geometry that is used to adjust the camera to a suitable viewing distance when rotating the network representation. The laser-pointer can thereby be used to explore local connectivity from the distance by high/low-lighting individual network neighborhoods.

In the detail perspective, the user has the option to ``fly'' through the network using free flying controls for immersive exploration. 
Input on the D-pad directly controls the user's position relative to his or her current viewing vector (pressing forward/backward on the D-pad lets users fly to/from where they are looking; pressing left/right allows the user to strafe perpendicular to the viewing vector of the VR headset).
Since 6 degrees of freedom (DOF) navigation in VR can cause discomfort for some users \cite{usoh1999walking, langbehn2018evaluation}, our system also offers a second navigation mode in the detail perspective: by selecting a hovered edge or node with the trigger button, an automatic flight transition to the selected graph element is triggered. The velocity is thereby increased in comparison to the free flying controls to enable faster navigation within the graph. Arrival at a node/edge is smoothed with an ease-out curve with the aim to further reduce motion sickness. Using this mode, the user can follow paths in the network from a first person perspective.

\textbf{Switching between overview and detail:}
To switch to the detail perspective from the overview, users must select a node of interest as their detail-exploration starting position.
The camera representing the detail perspective thus moves to the selected entry point.
By confirming their selection on the node containing the camera,
the user instantly teleports to the selected node position.
As teleportation may have a negative effect on the user's orientation when changing both the position \emph{and} the rotation after a teleportation \cite{moghadam2018scene}, we only translate the user's position, but keep the orientation of the user's starting perspective.
To switch back to the overview perspective, the user must simply click on the green overview-indicator arrow in the top of the view -- thus sparing users from tracking down their initial overview position before being able to switch perspectives.
Special consideration is thereby taken to avoid disorientation: when switching back to the overview perspective, the entire graph is placed directly in-front of the user's current viewing vector, on a height that matches the user's head's position in space.
This way, the user is able to instantly see the graph at eye-level when teleporting back to the overview, thus sparing them from unpleasant viewing angles and from searching the location of the graph.

Aside from rotation, free and automatic flying, and teleportation, our system also supports room-scale navigation, i.e., depending on the scale of the depicted network and the size of the allotted physical VR interaction space, users are able to physically walk around within the network, simultaneously changing their relative position and orientation in the VR scene.



\begin{figure}[t]
    \centering
    \includegraphics[width=\linewidth]{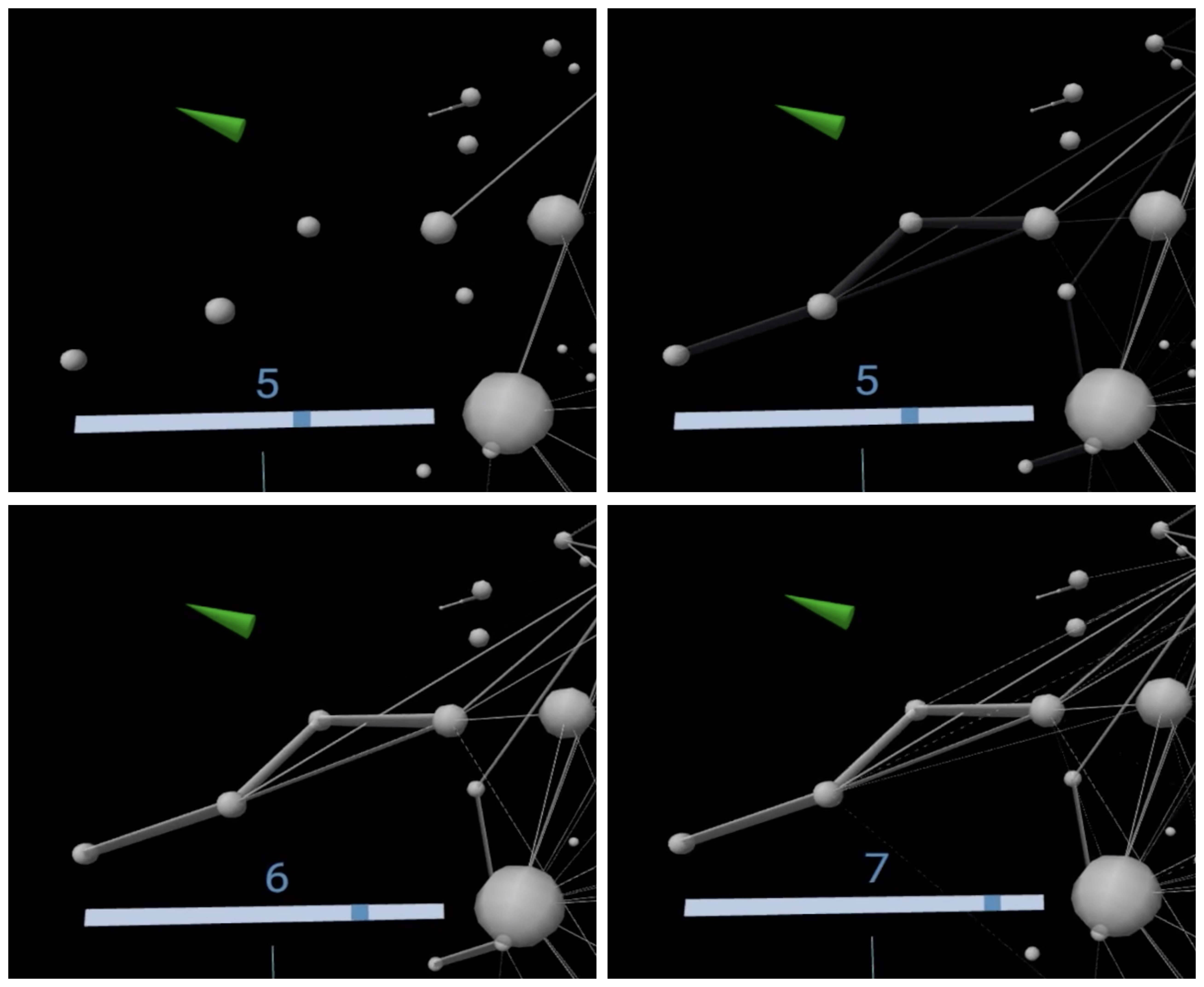}
    \caption{A sequence captured from a temporal transition between three time frames. The time-bar on top of the laser-pointer indicates the current time step.
    The top right image shows the transitional state between steps 5 and 6. The images are captured in detail perspective. The green arrow points to the position of the overview camera.
    }
    \label{fig:time}
\end{figure}

\textbf{Temporal Navigation: }The final navigation modality allows users to scroll through the temporal evolution of a dynamic network
, both from an overview and a detail perspective. Temporal navigation is carried out via the D-pad's horizontal input axis while pressing the input-modification button.
A time axis will appear on top of the laser pointer, indicating the current time step in relation to the time axis (see \autoref{fig:time}).
When the time frame changes, the nodes and edges of the current time step fade in, while graph elements not present in the current step are faded out, as depicted in first two images in the sequence of \autoref{fig:time}. The smooth fading thereby facilitates observation of the temporal change.
The layout of the graph (i.e., node and edge positions) is kept static to facilitate the comparison between different time steps and to avoid disorientation of the user.
The temporal navigation mode also lends itself to filtering graph elements by sliding through discrete attribute ranges instead of through time. Additionally, this navigation mode can be used for the comparison of two different (but related) networks by navigating (switching) from one set of nodes/edges to the other one.



\subsection{Implementation}
The system is implemented as a client-only web application, i.e., no server-side scripts need to be executed at runtime. The code is written completely in JavaScript using three.js~\cite{threejs} and A-Frame~\cite{aframe} -- a layer on top of three.js that supports browser based VR applications by handling the stereoscopic rendering as well as controller and head-tracking input from VR hardware.
The foundation of our system is based on an open source library for viewing graphs in VR~\cite{forcegraph}. This library handles the loading, layout, and rendering of the graph, and offers simple navigation via flying by mouse and keyboard outside of VR. In VR, the core does not offer any navigation or interaction capabilities (only looking and pointing via the headset's viewing vector). In order to support the navigation and interaction methods that form part of our contribution, we thus extended this core to accept input via VR controllers (specifically, the HTC Vive). The interface between the VR application (i.e., the web-browser) and the VR hardware is handled by SteamVR~\cite{steamvr} that natively supports the HTC Vive.

Rendering large node-link diagrams in 3D can have a big impact on performance as all graph elements are rendered as geometries. To improve performance, edges can alternatively be rendered as simple lines. However, this reduces their visibility, makes them harder to select, and removes the support for encoding edge weights in the girth.
To avoid such compromises in usability and readability, we further extended the core library's rendering capabilities to support instancing of node and edge geometry via three.js \textit{InstancedBufferGeometry}. The improvement almost doubles the achieved frame rate on the largest graph of the performance evaluation (2000 nodes, 6000 edges \autoref{se:eval_quantitative}), from 13 to 22 frames per second.

Special consideration also is needed for the implementation of the dual camera perspectives and associated navigation and interaction modalities. The camera orientation, and VR controls are child objects of a parent container, the \textit{active rig} that represents the user's position in the VR scene.
To maintain two perspectives, a second rig, the \textit{passive rig}, serves as a placeholder, storing the other perspective's position and orientation. When switching between perspectives, position and orientation of the two rigs are simply interchanged.

\section{Evaluation}\label{se:eval}

We evaluate our approach by means of a quantitative and qualitative evaluation. The former is carried out as a performance analysis of the tool, in terms of a frame rate benchmark with networks of increasing size (see \autoref{se:eval_quantitative}). The latter is carried out through a case study with two domain experts who were asked to perform a walk-through in a real healthcare dataset to evaluate how our design choices reflect on their use of the system (see \autoref{se:eval_casestudy}). Both experiments are run on the same test setup, on a desktop PC equipped with an AMD Threadripper 1900-X CPU with 32GB of RAM and a Geforce GTX 1080 Ti graphics card. The headset used is an HTC Vive head-mounted display and controllers. 

\subsection{Performance and Frame Rate}\label{se:eval_quantitative}

\begin{figure}[t]
    \centering
    \includegraphics[width=0.95\linewidth]{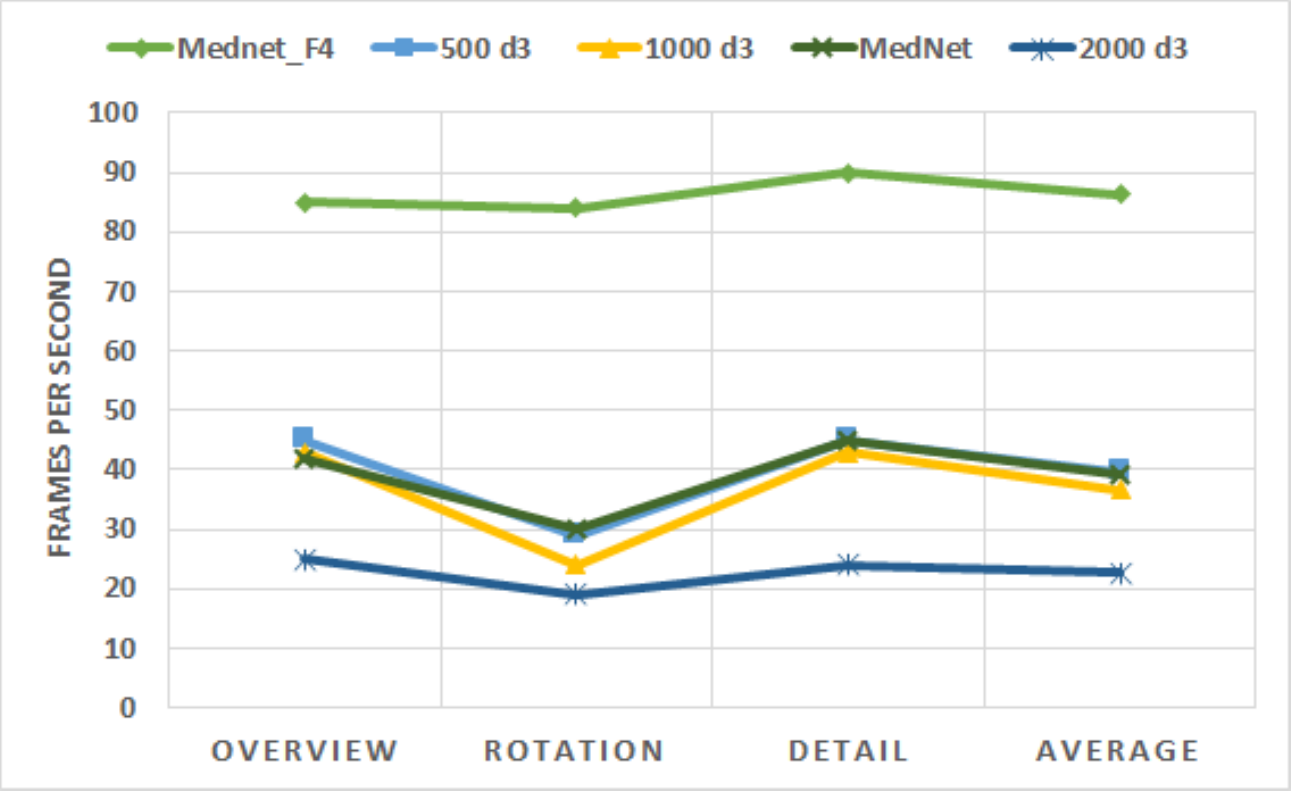}
    \caption{System performance comparison. On the y-axis the average frames per second; on the x-axis the perspective in which the measurement was taken.}
    \label{fig:chart_fps}
\end{figure}

To enable immersion and avoid motion sickness, it is necessary to render stereoscopic images with a fluid frame rate. 
For that matter, SteamVR automatically stops sending updates to the headset when performance drops below 20 frames per second (fps).
To assess the performance of our system, we simulate a typical use case scenario on increasingly large graphs and record the frame rate in three configurations: a static \emph{overview}, \emph{rotation} of the network in the overview, and navigation in the \emph{detail} perspective.
As a benchmark, we created three random graphs (using the Erd\"os-Reyny~\cite{erd6s1959random} model), with 500, 1000, and 2000 nodes, and an average node degree of 3 (i.e., with 1500, 3000, and 6000 edges respectively). Additionally, we use 2 real-world datasets that include the one used in our case study: \emph{MedNet F4} with 199 nodes and 593 edges (see \autoref{se:eval_casestudy}) and \emph{MedNet} with 692 nodes and 3047 edges, for a total of five graphs. The results are reported in \autoref{fig:chart_fps}. The chart shows that with the smallest graph the average fps rate is about 85, with no major difference on the two perspectives. As the size increases, the average frame rate drops down to 35-40 fps with a low of about 22 fps for the largest graph. The rotation in the overview perspective requires the most computing power, since the entire graph is on screen while being animated. In the detail perspective instead, only a portion of the graph is visible thus reducing the load on the graphics hardware. It is worth mentioning that on real-world datasets, performance was satisfactory: the \textit{MedNet F4} graph, used during the case study (see \autoref{se:eval_casestudy}) rendered with $\sim$85 fps for smooth navigation and interaction. When pushing the system with the last graph (2000 nodes, 6000 edges), the large number of elements on screen had a significant impact on rendering, with frequent stutter especially during ray-casting with the laser-pointer. 

\begin{figure}[t]
    \centering
    \includegraphics[width=\linewidth]{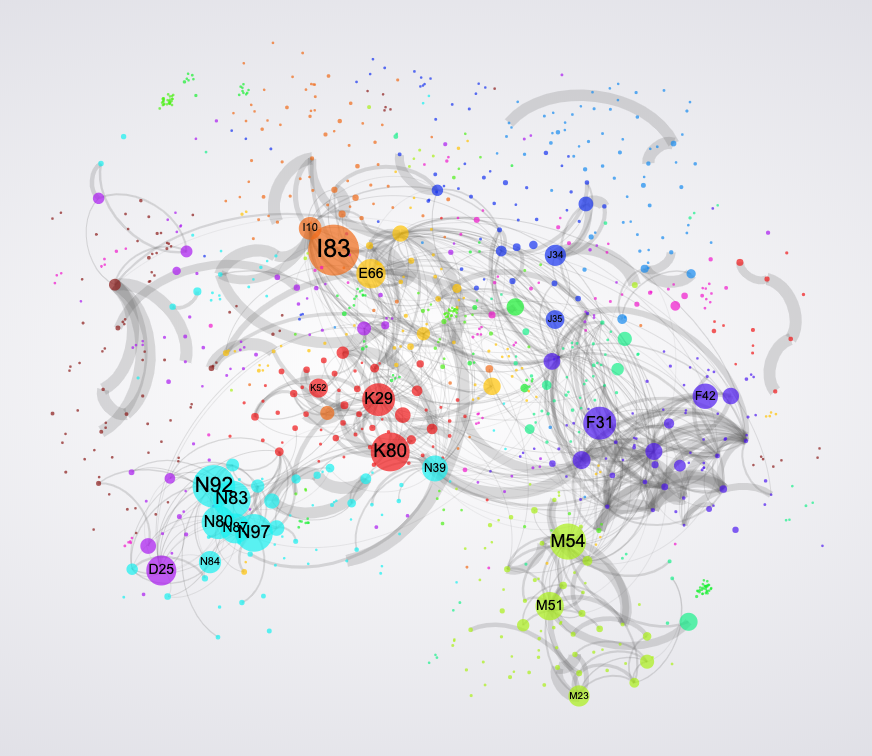}
    \caption{The semi-automatic 2D layout of the co-morbidity graph created by our domain experts.}
    \label{fig:2d_graph}
\end{figure}

\subsection{Case Study}\label{se:eval_casestudy}

The goal of the case study was to gain first realistic insights about the usefulness of the two network perspectives. We were interested to find out whether users see the necessity to switch their perspective and which kinds of discoveries they make in either perspective. This helps us to provide a first characterization of the potential mutual benefits of the two perspectives for immersive network analytics. In addition, we asked users to report symptoms of motion sickness, disorientation, or other navigation problems to identify potential issues after switching views.

For the case study, we asked a female data scientist and a male physicist to present and explore a co-morbidity network~\cite{chmiel2014spreading} that they are currently investigating in the course of their research. They are already working with the dataset for several months (data scientist) to years (physicist) to analyze patterns of diseases that usually occur together and to compare these patterns across different populations, such as male and female patients. In their standard workflow, they use Gephi~\cite{bastian2009gephi} for generating 2D visualizations of the entire graph. These visualizations are printed and presented to collaborating medical experts for analysis. The giant component of the network, which was also used for this case study, contains 199 nodes and 593 edges.
The domain experts consider their current approach using static 2D graphs as partially insufficient as it is not possible to interactively dig into the network to get more details. They also reported difficulties finding a good graph layout, where groups of strongly connected diseases are clearly visible.
To try to overcome those complications, they currently analyze and present their network by computing an initial layout using a force-directed layout and subsequently manually moving certain nodes to separate (known) clusters. Additionally, they only render a subset of all edges to reduce visual clutter. The semi-automatic 2D layout is shown in \autoref{fig:2d_graph} while the (fully-automatic) 3D layout of the same network is in \autoref{fig:overview}. 

After an initial interview about their current workflow, we gave a short introduction to the VR setup and the interaction techniques without revealing what is novel about the navigation approach. We then asked the users to walk through their data set and to orally present the contents as if they would present them to their medical partners, as well as to comment on the usability and appropriateness of the VR setup. We recorded the VR sessions, and transcribed the videos afterwards. We coded all user findings in the transcripts, and also coded if the finding was already known or unexpected, as well as whether it was made in the overview or in the detail perspective. In addition, we extracted all comments and suggestions about the VR setup. The data scientist spent 33 minutes, and the physicist spent seven minutes in the VR environment.

Both users obeserved and reported a number of known facts about the network during the session. These reported facts covered \emph{clusters} (such as a clearly separate cluster of cancers or the very central cluster of mental diseases, see \autoref{fig:overview}-B), \emph{hub nodes} (such as breast cancer or depressive episodes, see \autoref{fig:overview}-A), \emph{paths} between nodes, and isolated \emph{sub-networks} (such as teeth- or pregnancy-related problems, see \autoref{fig:overview}-C). Most of these known facts were reported while exploring the graph in the overview. This is not surprising, since the users knew the dataset before and are also used to see it laid out in 2D. As the physicist explained, the overview is similar to their well-known 2D view of the graph: \emph{``Overview also works in 2D. If I look at it now, it does not look too different than if looking at an image.''} The data scientist also appreciated that she could get an overview first: \emph{``I can see many many interesting things and then when I decide what will be the focus, then I can go into it and look at it, for example, from perspective of overweight} [a node in the network].''

Both users switched to the detail perspective to explore the network from various nodes at least once in the course of the study. Both users found this perspective particularly interesting. The physicist explained that he \emph{``can browse these hubs more easily than in 2D''}. Indeed, known facts reported while being in the detail perspective were about connections to single nodes. For instance, the data scientist explored the nodes directly connected to obesity: \emph{``I can see what will happen when I eat too much. So I will get diabetes [...], I will be depressed. Of course. Ok, I will have sleep disorder. So these things are quite known. It’s super interesting to see them from this point of view.''}

While the physicist did not report any unexpected findings, the data scientist reported a few previously unknown discoveries. Around half of these discoveries thereby were made in the detail perspective. For instance, she found that some cancers were separated from the main cancer cluster and that endocrine diseases are quite separated from other diseases. From the overview perspective, she was surprised to see that mental diseases seem to be the most connected cluster in the network and that some diseases are far away from other diseases, like tonsillitis.

Both users reported they could not see all edges and their encoded weights when observing the graph from a node point of view in the detail perspective.
Also, they both expressed the wish to be able to easily traverse the network specifically along directed edges in the network to be able to follow a \emph{``patient `career' from the perspective of the diseases.''} What both users appreciated was the ability to easily obtain the labels of the nodes. The data scientist explained that only in VR, it is possible to render all the edges and still to be able to read all the associated labels.

From our case study it seems that users are more prone to receive symptoms of motion sickness in the detail perspective than in the overview. During her case study, the data scientist reported three times that her \emph{``brain is confused''} after moving in the detail perspective. Twice, this happened after free flying, once after an animated teleportation to the neighboring node. She did not report any motion sickness symptoms when rotating the network in the overview perspective. It should be noted that the overall well-being and engagement during the study appeared to be very high, since the data scientist was very surprised to hear that she spent over half an hour exploring her graph in VR. After the study, she reported that to her it seemed that her session had only lasted about five minutes.

\subsection{Lessons Learned and System Limitations}\label{se:eval_lessons}

Our case study has shown the potential of combining a well-known overview of a network with a first person detail perspective, which is only possible in an immersive environment. The overview thereby represents a known reference frame, while the detail provides a new, egocentric perspective into the network.
We found that the detail perspective facilitates the analysis of local neighborhoods, especially when the nodes of interest lie within a dense part of the network.
Users may discover unexpected information in known networks from such an unknown perspective, such as nodes separated from their main cluster or unexpected connections between nodes. User feedback also indicates room for improvement for such detailed network perspectives: the weight of the edges should be better visible and flying to adjacent nodes can lead to motion sickness symptoms.


The presented case study has to be considered as early feedback. The limitations are, besides the limited number of users and the single data set, that both users knew the data set before. However, we believe that this is a common scenario in the case of network analytics, where VR will most likely represent a complementary tool rather than the only way of how to visually inspect a network. To evaluate the role of VR in data analysts' workflows and to improve a fluid transition between desktop and VR analysis, longer-term field studies will be necessary. In the future, it will also be necessary to formally compare the types of insights users gain from exploring networks solely from a detail, solely from an overview, or from a switchable overview\&detail navigation in a controlled user study. Our expectation is that being able to switch between those two views will strengthen the users' understanding of the network topology. The fact that our case study users switched between those views and reported facts both, from the overview and the detail perspective, are first indications of the strength of the approach.

\section{Conclusion and Future Work}\label{se:conclusion}

In this paper, we presented a new approach to facilitate network analytics by combining well-known exocentric overviews with an immersive detail perspective in VR. This immersive perspective provides insights into the network that are not possible in a conventional desktop setting. Our case study has shown that users find this new detail perspective interesting and that it supports them in gaining new, detailed insights into known data. The case study also indicated that the initial overview is important to stay oriented. Our web-based implementation thereby enables fluid exploration of networks wherever an HMD-based VR setup is available, which is an important step to make immersive network analytics accessible to a broad spectrum of users. Our results open up new exciting research opportunities for graph drawing in VR -- especially in the context of immersive detail inspection of networks.


\section*{Acknowledgments}
This work was supported by the FFG Project 857136 at CSH Vienna, as well as the Research Cluster ``Smart Communities and Technologies (Smart CT)'' at TU Wien.

\bibliographystyle{IEEEtran}

\bibliography{bibliography}

\end{document}